\documentclass[letter,scriptaddress,twocolumn, prl,showkeys]{revtex4}

	\usepackage{amsmath}
	\usepackage{makeidx}
	\usepackage{amsfonts}
	\usepackage[ansinew]{inputenc}
	\usepackage[usenames,dvipsnames]{pstricks}
	\usepackage{subfigure}
	\usepackage{epsfig}
	\usepackage{pst-grad} 
	\usepackage{pst-plot} 
	\usepackage[colorlinks,hyperindex]{hyperref}
	\hypersetup
	{
		colorlinks,%
		citecolor=black,%
		linkcolor=black,%
		urlcolor=black,%
	}

\makeindex

\begin{document}

\title{Orientational ordering in crumpled elastic sheets}

\author{Anne Dominique Cambou}
\email{dominique@physics.umass.edu}
\author{Narayanan Menon}
\email{menon@physics.umass.edu}

\affiliation{Department of Physics, University of Massachusetts, Amherst, MA 01003-3720}

\date{\today}

\begin{abstract}
We report an experimental study of the development of orientational order in a crumpled sheet, with a particular focus on the role played by the geometry of confinement.  Our experiments are performed on elastomeric sheets immersed in a fluid, so that the effects of plasticity and friction are suppressed. When the sheet is crumpled either axially or radially within a cylinder, we find that the sheet aligns with the flat or the curved wall, depending on the aspect ratio of the cylinder.  Nematic correlations develop between the normals of the sheets at relatively low volume fractions and the crumpled object has large density fluctuations corresponding to the stacking of parallel sheets. The aligning effect of the wall breaks symmetry and selects the direction of ordering.
\end{abstract}


\maketitle

At first sight, a thin sheet crumpled into a ball appears to be a very spatially disordered object. When crushed within a shrinking container a sheet collapses into a complicated 3-dimensional labyrinth of nearly flat facets bounded by ridges or creases. The geometry of a piece of paper balled within your fist has indeed been found to be complex, statistically variable, and not ordered in any obvious fashion ~\cite{gomes, blair, balankin, cambou, linpressure}. However, another limit of this process leads to a different intuition. Imagine crushing a can underfoot, or squashing a crumpled object in a trash compactor: as the thin sheet is confined to volume fractions near unity, intuition suggests that the facets of the crumpled object must lie flat, in parallel stacks. One wouldn't expect this arrangement to be ordered like the folds of an accordion, but nonetheless, the arrangement could have underlying long-range orientational order, with the normals to most facets aligning parallel to the confining direction.

In this article, we discuss evidence that spatial ordering underlies the arrangement of material in a crumpled object.  The examples above supply the intuition that parallel stacks are a likely outcome at high density if you squash the object flat. What if one confines the sheet to high density, but not with flat confining surfaces? One of the principal concerns of this article is whether stacking is merely an outcome of conforming to the crushing boundary or if it is a geometric inevitability in trying to pack a sheet into any volume at very high density. The experiments we describe here study the development of orientational order in the sheet as a function of both the external boundary's shape and the degree of confinement, that is, the aspect ratio as well as the volume fraction. 

Considerations of ordering in confined flexible objects have been better fleshed out in the situation of a 1-dimensional object confined in a 2-dimensional area. Experiments \cite{deboeuf, donato, bayart} and simulations \cite{stoop} show that a 1D curve confined within a circle forms a series of loops that conform to the circular boundary. Simulations that use simplified models of bending ~\cite{aristoff}, and mean field models of elastica ~\cite{katzav} have shown the existence of a sharply-defined nematic transition where long-range orientational order sets in. For elastic rods of a finite thickness, ~\cite{boue} and ~\cite{katzav} predict  a jamming transition further within the nematic phase. 

Confining a 2D sheet into a 3D volume, however, is qualitatively different. Rather than coiling smoothly, even a purely elastic sheet will localize deformation ~\cite{cerda, wittenRMP, lobkovsky} and form sharp ridges and vertices under confinement. Furthermore, the constraints of self-avoidance are more severe in this case. Experimentally, the conformation of a curve in a plane is much easier to visualize than a crumpled membrane in a volume, and therefore the 2D analogue has been more studied. However, x-ray tomography has been used to study the internal conformation of aluminum foil crumpled into a ball~\cite{lin},~\cite{cambou}. Lin \textit{et al.} studied 2D slices of a crumpled ball and found ordered domains whose size increased with compaction. In our own work~\cite{cambou}, we made 3D analyses of the orientation of the sheet within the volume of the ball. The orientation was found to be isotropic at all radial locations within the ball, however, we found evidence of stacks of nematically oriented layers forming preferentially near the boundary and growing with increasing compression. The shape of the boundary was not varied in either of these studies, and a spherical boundary with gaussian curvature is not a favorable geometry for the development of nematic stacks in a thin sheet that does not want to stretch to conform to the boundary.

We have therefore designed an experiment in which we confine a sheet within a cylindrical container and vary the type of confinement by altering either the radius or the height, thus increasing the relative amounts of flat boundary and of boundary with mean curvature.  In addition to the geometric constraints of self-avoidance, material properties such as friction and plasticity hinder relaxation towards a lower energy state. Simulations \cite{tallinen} and experiments \cite{aharoni} show that in the absence of friction and plasticity, the folds and vertices are mobile, thus allowing the membrane to assume lower energy configurations. In the current experiments we eliminate plastic effects by using elastomeric sheets so that any deformation is reversible~\cite{aharoni}. The sheet is immersed in a fluid, so that different lamina of the sheet can slide by each other without friction.  Finally, we visualize the sheet by optical tomography; unlike the slower x-ray technique, optical scans allow us to observe the time-dependent evolution of the conformation. 

\begin{figure} [htbp]
	\begin{center}
		\includegraphics[width=0.5\textwidth]{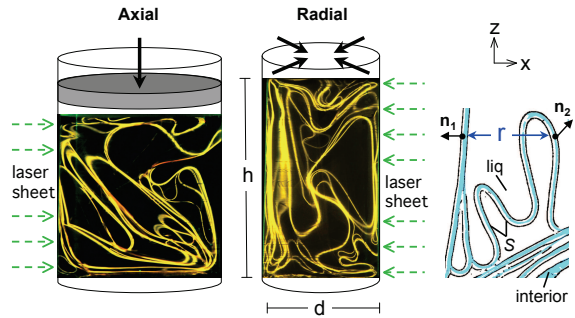} 
		\caption{Crushing geometry. A schematic view with images of the central cross-section in the X-Z plane of the cylinder illuminated by a laser sheet. Left: axial geometry where the height can be increased or decreased between h $\approx$ 4cm to 1cm. Middle: radial geometry where the diameter, d, can be reduced from $\approx$ 4cm to 2cm. Right: Surface normals, $\hat{n}_i$, are determined with an edge detector that divides the images into three distinct regions:points at the surface  (S, dark),  points inside the film (interior, light), (both with mass  $m = 1$) and points in the liquid  (liq, white) ($m=0$).}
		\label{fig:setup} 
	\end{center}
\end{figure}

We make polydimethylsiloxane (PDMS) sheets using a 10:1 mixing ratio of Sylgard 184 spread onto a silanized glass substrate and cured in an oven at $70 \deg C$ overnight. Squares of area $A_{sheet}=150$ to $200~cm^2$ are cut out of the PDMS and infused with a hydrophilic dye (rhodamine B). The sheet is introduced into the experimental cell where it is submerged in dodecane, which closely matches the index of refraction, $n \simeq 1.4$.  Dodecane swells the sheet by about $50\%$ to a final thickness of  $t \simeq 37.5 \mu m$. 

The PDMS sheet is initially loosely confined within a cylindrical volume of height $h$ and diameter $d$ defined by transparent teflon walls (Fig.\ref{fig:setup}). These walls maintain the prescribed cylindrical shape even when confining the crumpled object. The sheet is crumpled either axially, by pushing down with a flat piston, or radially, by reducing the diameter of the cylinder. These two independent ways of crushing the film allow us to vary the aspect ratio $\mathcal{A} = \frac{h}{(d/2)}$ as well as the volume fraction $\phi = \frac{t A_{sheet}}{V}$, where the confining volume, $V = \pi (d/2)^2 h$.

We illuminate a cross-section in the X-Z plane in the interior of the crumpled sheet  (Fig. \ref{fig:setup}) with a $532nm$ laser sheet of thickness $\approx$ $1mm$. The fluorescence of the sheet is imaged with a Nikon D80 camera through a filter that eliminates the illumination line. The thickness of a sheet normal to the image plane is $\approx$ 3 pixels. Here we work with 2D slices taken at the central cross-section of the cylinder. To determine the locations of surface points and find surface normals $\hat{n}$, a canny edge-detection is applied to the grey-scale image that has been smoothed with a mean filter. 

A typical experimental protocol involves crumpling, axially for instance, by reducing the confining height $h$ from the initial cylinder height in discrete steps towards a maximum volume fraction $\phi$, and then reversing the motion. Each finite step is taken in 5 secs, followed by waiting a time $\tau_w$ for the sheet to relax.  Most of the reconfiguration of the sheet is completed over a relaxation time, $\simeq 10$ $min$, however small local motions, presumably triggered by ambient noise, persist for hours. The quantities we study here are insensitive to waiting time if  $\tau_w >$ 30 secs.



To study the effect of the confining boundary on the development of alignment, we determine the orientation of normals, $\hat{n}$, to the sheet at all locations and construct the scalar product with the normal ($\hat{z}$ or $\hat{x}$) to the top or side walls. We note that $\hat{n}$ is only the 2D projection in the XZ-plane of the true 3D normal to the sheet, so that completely random alignment of the 3D vector leads to an average value of $<\hat{n}\cdot \hat{z}> =2/\pi$.

\begin{figure}[htbp]
\begin{center}
\includegraphics[width=0.5\textwidth]{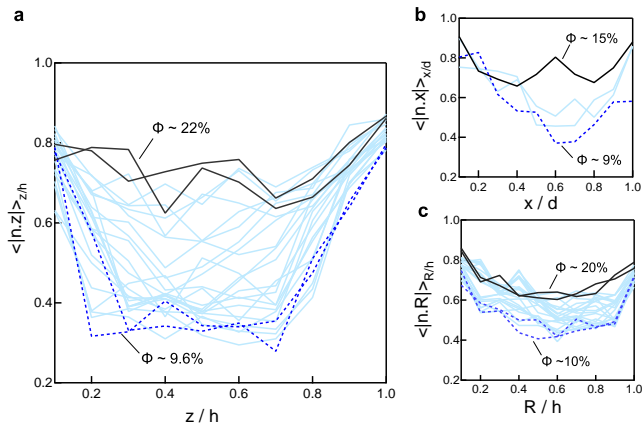} 
\caption{Alignment with boundary. (A) We display $<|\hat{n}\cdot \hat{z}|>_{z/h}$ as a function of distance $z$ from the moving top boundary. When normalized by the height $h$, $z/h=0$ is at the piston and $z/h=1$ at the bottom of the cylinder. We show a single run of an axial crumpling experiment going from $\phi \simeq 9.6 \%$ (dark blue dashes) to $\phi  \simeq 22 \% $ (black curve) and back with a number of intermediate volume fractions (light blue). The alignment is greatest at the top and bottom of the cylinder and grows inward as $\phi$ increases. Similar behaviour is seen in the other crumpling geometries, (B) radial,  where $x = 0,d$ at the left and right cylinder walls respectively, and (C) hemispherical top boundary  where $R = 0,h$ at the piston and bottom of the cylinder respectively.}
\label{fig:nzxR}
\end{center}
\end{figure}

In Figure \ref{fig:nzxR} A, we show $<|\hat{n}\cdot \hat{z}|>_{z}$ averaged over horizontal position, as a function of the distance from the top confining wall $z/h$ in an axial crumpling experiment. Even in the initial, least-crumpled, state, there is considerable alignment in the vertical direction near the bottom and top walls, but the alignment decays rapidly in the bulk of the system. As we compress with the flat piston, alignment increases everywhere in the bulk of the cell, until the alignment is extremely high everywhere in the most compressed state achieved in this run. Subsequent release of the compression reverses this trend. A similar progression is shown in Fig.\ref{fig:nzxR}B for radial crumpling, where $<|\hat{n}\cdot \hat{x}|>_{x}$, the alignment in the radial direction increases.  Boundary alignment also increases for the more complicated case in Fig.\ref{fig:nzxR}C, where the compressing piston is replaced by a hemispherical cap. In all these geometries, the tendency to align is much stronger than in the case of crumpling in a sphere~\cite{cambou} where all boundaries have gaussian curvature and defeat the tendency to conform to the boundary.

\begin{figure}[htbp]
\begin{center}
\includegraphics[width=0.5\textwidth]{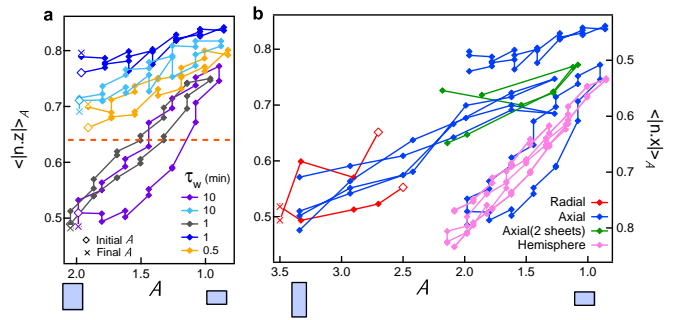}
\caption{Spatially averaged alignment. (A) We show $|\hat{n} \cdot \hat{z}|$ averaged over the entire cross-section of an axially crumpled sheet. Data are shown for several instances of crumpling and uncrumpling cycles with the initial and final data points being indicated with an open diamond and a cross, respectively. The rectangles below the x-axis represent the aspect ratio $\mathcal{A}$ at the extremes of crumpling and uncrumpling. Orientation along the z-axis grows as $\mathcal{A}$ decreases. The value of isotropic orientation in 2D is shown by the dashed line. Even though there is no plastic deformation in the sheet, it maintains memory of its initial configuration as it returns to a similar average alignment. The waiting time, $\tau_w$, appears to have little effect on the evolution of orientation.  (B) Average alignment for varying confinement geometries, radial, axial, axial with two sheets, and hemisphere (red, blue, green, and pink respectively). The curves indicate increasing alignment with the majority boundary. The right axis shows that as alignment increases in one direction, it decreases in the other. }
\label{fig:align_aspect} 
\end{center}
\end{figure}

The orientation of the alignment is revealed to be governed by the dominant boundary. In a cylindrical geometry, the relative surface area of the cylindrical and flat (top and bottom) boundaries is given by the aspect ratio  $\mathcal{A} = \frac{h}{(d/2)}$.  In Figure \ref{fig:align_aspect}, we explore the evolution of the alignment averaged over the entire cross-section as a function of aspect ratio. In Figure \ref{fig:align_aspect}A, we display $<|\hat{n}\cdot \hat{z}|>_{\mathcal{A}}$ versus $\mathcal{A}$ for several axial crumpling experiments. We point out a few distinct features of the progression:(i) The increase in alignment with increase in aspect ratio is robust. (ii) The initial alignment varies considerably from run to run, and one complete loop  of crumpling and uncrumpling does not take the conformation very far from the initial level of alignment. (iii) At each aspect ratio we show two data symbols representing images taken immediately after a compression step, and just before the next step is taken. None of the averaged quantities we show are sensitive to the waiting time $\tau_{w}$ between the steps, even though subtle reconfigurations occur over longer times. 

Both small and large values of $\mathcal{A}$ might be expected to create alignment by conforming to the developable flat and cylindrical surfaces by disordered accordion- and scroll-like motifs, respectively. In Figure \ref{fig:align_aspect}B, we confirm this expectation by combining results from both radial and axial crumpling experiments, spanning both sides of $\mathcal{A}=2$, which denotes a square cross-section. At small $\mathcal{A}$, the normals align radially, and at large $\mathcal{A}$, they align axially.

Thus far, we have only considered the aligning effect of the boundaries in Figures \ref{fig:nzxR} and \ref{fig:align_aspect}. However, there are two further questions one may ask. The first is whether orientation is only a boundary effect, with a finite penetration depth, or whether there is a tendency to orient in bulk as the volume fraction is increased. In this latter scenario, the boundary merely breaks symmetry and chooses the direction of orientation.

To assess the development of orientational order in the bulk, we consider normal-normal correlations($<\hat{n} \cdot \hat{n}>$), rather than correlations between the sheet normal and the normal to the boundary (e.g. $<|\hat{n} \cdot \hat{z}|>$). For measuring local ordering, we correlate pairs of surface normals separated by a distance $r$ (see Fig. \ref{fig:setup}). We divide the cross-sectional images into three regions: fluid, where $m(\vec{x})=0$;  bulk, where $m(\vec{x})=1$; and surface, where $m(\vec{x}) = 1$ and $\vec{x}\: \epsilon \: S$ (see Fig. \ref{fig:setup}).


We define the normal-normal correlation function as
\begin{equation}
\label{E:adj_nncorrfunc}
C_{n}(r) = \sum\limits_{i=0}^{N_{S}} \frac{|\hat{n}_{i} \cdot \hat{n}(\vec{x_{i}} \pm \hat{n}_{i}r)|}{1+N^\pm_{liq}} \;\;\; \bigg| 
\begin{array}{l}
 \vec{x}_{i}\; \epsilon \: S\\
\vec{x_{i}} \pm \hat{n}_{i}d\; \epsilon \: S\\
 \hat{n}(\vec{x}_{i}) = \hat{n}_{i}
\end{array}
\end{equation} 

Here the total number of surface points in each image is $N_{S}$ and the correlation function is normalized by the number of liquid points between each pair, $N_{liq}$.

The amount of layering is quantified by the density correlations at a distance, $r$, away from each point on the surface, $\vec{x}_{i}$ in the direction of the sheet normal , $\hat{n}_{i}$. 

\begin{equation}
\label{E:denscorrfunc}
C_{m}(r) = \frac{\sum\limits_{i=0}^{N_{S}} m(\vec{x}_{i} \pm \hat{n}_{i}r)}{2 N_{S}}\;\;\; \bigg| 
\begin{array}{l}
\vec{x_{i}} \; \epsilon \: S\\
  \hat{n}(\vec{x_{i}}) = \hat{n}_{i} 
\end{array}
\end{equation}

\begin{figure}[htbp]
\begin{center}
\includegraphics[width=0.5\textwidth]{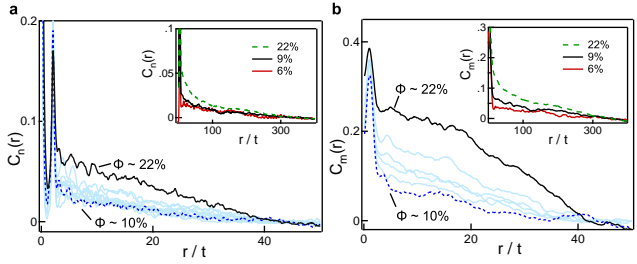} 
\caption{Correlation functions.  We show (A) the normal-normal orientational correlation function, $C_{n}(r)$, and B) the density correlation function in the normal direction, $C_{m}(r)$, for a sheet crumpled axially from $\phi \simeq 10 \%$ to $\phi \simeq 20 \%$ (dark blue and black curves respectively). The curves are normalized to decay to zero at large $r/t$. Both graphs show the correlations decaying more slowly as the sheet is confined to higher $\phi$. Insets: The corresponding correlation functions of a 2D cross-section of an aluminum sheet crumpled into a sphere~\cite{cambou}.}
\label{fig:nnmass}
\end{center}
\end{figure}

In Figure~\ref{fig:nnmass} we show for a cycle of axial crumpling and uncrumpling steps, the development of both orientational order, as monitored by  $C_{n}(r)$, and of density correlations in the normal direction, as monitored by $C_{n}(r)$. 
There is a robust trend for the density and orientation correlations to decay slower with increasing $\phi$, even at these rather low volume fractions. The corresponding graphs from our earlier experiments \cite{cambou} with aluminium foil crumpled into spheres are shown in the inset where the overall orientation of surface normals was found to be isotropic, unlike the aligned states achieved within a cylinder (cf. Fig \ref{fig:align_aspect}). However, the evolution with volume fraction is qualitatively similar, thus indicating a propensity to orient even in the absence of a boundary that promotes alignment.

\begin{figure}[htbp]
\begin{center}
\includegraphics[width=0.5\textwidth]{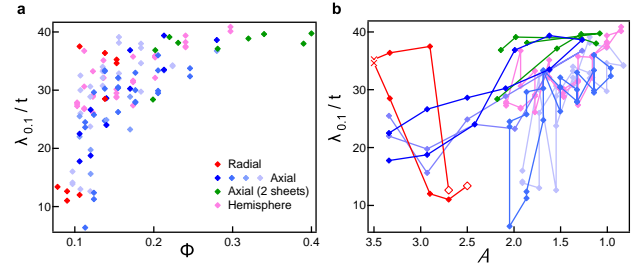}
\caption{Growth of orientational correlations.  A) The correlation length $\lambda_{0.1}$ versus  $\phi$ shows that the range of orientational order grows monotonically with the amount of crumpling in all our experiments. B) $\lambda_{0.1}$ versus $\mathcal{A}$. The open diamonds and crosses show the beginning and the end of the  radial crumpling sequence. Here $\mathcal{A}$ increases as $\phi$ increases. Volume fraction more strongly influences orientational ordering than does $\mathcal{A}$.}
\label{fig:avcn} 
\end{center}
\end{figure}

As either radial or axial crumpling proceeds, both the degree of confinement as well as the effect of the walls change. To assess how each of these factors contributes to the development of correlations,  we extract a  length scale $\lambda_{k}$ from the correlation function $C_{m}(r)$  by finding the distance $r$ at which this function decays to a fraction $k (0<k<1)$ of its peak height. A similar measure can be obtained from $C_{m}(r)$. In Fig. \ref{fig:avcn}, we show $\lambda_{k}$  for $k=0.1$ plotted against volume fraction $\phi$ as well as the aspect ratio $\mathcal{A}$.   Both variables clearly affect the development of orientational order, but it is clear that the effect of $\phi$ is stronger than that of $\mathcal{A}$. For example, $\overline{C_n}$ grows when radial crumpling pushes the aspect ratio below $\mathcal{A}=2$. However, for axial crumpling, $\overline{C_n}$ does not grow below $\mathcal{A}=2$, because the volume fraction is lowest here. This is the case even though alignment increases in this range, as shown in Figure \ref{fig:align_aspect}. The qualitative conclusion is that orientational order grows with volume fraction, even if there is no effect of boundary alignment (see also Fig.\ref{fig:nnmass} inset).

%

To summarize, we observe that both cylindrical and flat walls promote alignment of a crumpled sheet.  However, it appears that this is not merely a boundary effect. The process of crumpling leads to the development of orientational order at remarkably low volume fractions of between $8\%$ and $20\%$.  The principal role of the walls, as quantified by the aspect ratio, is to serve as the agent of symmetry-breaking and choose a preferred orientation for the sheet normals. The orientational order has two related aspects: one of these is that the sheet develops nematic order, and the other is that the sheets stack closely. The normals do not form a dilute, dispersed nematic, but are inhomogeneously clustered in stacks. We are not aware of any other state of matter with closely analogous properties, which presumably arises from the combined nonlocal demands of self avoidance  and of connectivity.  In our current experiments, we have not reached a regime of  long-range order, but it would clearly be important to study how both these aspects of ordering progress at higher densities, and whether either or both of them lead to true phase transitions, analogous to those predicted ~\cite{boue} for the lower dimensional system of an elastic rod confined in a 2D space. 

\begin{acknowledgements}
We gratefully acknowledge support from NSF DMR 12-0778  and 09-07245, NSF-MRSEC on Polymers at UMass Amherst DMR 0820506. We thank Tom Witten, Michael Bartlett, and Dayong Chen for helpful conversations. 

\end{acknowledgements}


\end{document}